\documentstyle[aps,preprint,tighten]{revtex}

\newcommand{\beq}{\begin{eqnarray}}
\newcommand{\eeq}{\end{eqnarray}}

\newcommand{\la}{\langle}
\newcommand{\ra}{\rangle}
\newcommand{\qc}{\la \bar{q}q \ra}
\newcommand{\uc}{\la \bar{u}u \ra}
\newcommand{\dc}{\la \bar{d}d \ra}
\newcommand{\s}{\la \bar{s}s \ra}
\newcommand{\gc}{\la {\alpha_s \over \pi} G^2 \ra}
\newcommand{\mqc}{\la \bar{q} g_s \sigma \cdot G q \ra}
\newcommand{\muc}{\la \bar{u} g_s \sigma \cdot G u \ra}
\newcommand{\mdc}{\la \bar{d} g_s \sigma \cdot G d \ra}
\newcommand{\msc}{\la \bar{s} g_s \sigma \cdot G s \ra}
\newcommand{\qsl}{\rlap{/}{q}}
\newcommand{\psl}{\rlap{/}{P}}
\newcommand{\gknl}{g_{KN\Lambda}}
\newcommand{\gkns}{g_{KN\Sigma}}

\begin{document}
\draft
%
%------------------------- Title ------------------------------------%
%--------------------------------------------------------------------%
%\preprint{\vbox{ \hfill SNUTP 95--082}}
\title{Kaon-baryon coupling constants in the QCD sum rule approach}
\author{Seungho Choe\thanks{E-mail : schoe@hirohe.hepl.hiroshima-u.ac.jp}}
\address{Department of Physics, Hiroshima University, Higashi-Hiroshima
739--8526, Japan }
%\date{}
\maketitle
\begin{abstract}
We improve our previous QCD sum rule calculation on $\gknl$ and
$\gkns$ coupling constants by including the contributions from
higher dimensional condensates, $\mqc$ and $\qc \gc$, in the OPE.
It is found that the contribution of these condensates is
non-negligible compared to that of the quark condensates. Using a
best-fit analysis we find $|\gknl|$ = 2.49 $\pm$ 1.25 and
 $|\gkns|$ =  0.395 $\pm$ 0.377.
\end{abstract}
\vspace{1cm}
%
%------------------------ Text --------------------------------------%
%--------------------------------------------------------------------%
\section{Introduction}

To understand kaon-nuclear physics, it is important to know
the hadronic coupling strengths involving the kaons.
Among them, $\gknl$ and $\gkns$ are the most relevant coupling
constants.
In contrast to $g_{\pi NN}$, however, the determination of these kaon
couplings has some difficulties both in the experimental side
and in the theoretical side, e.g. see \cite{chyc96}.

Among other  theoretical approaches,
QCD sum rule method \cite{svz79,rry85,narison89} has been used
to extract these kaon couplings.
However, compared to the large number  of works devoted to
$g_{\pi NN}$,  there have been only few QCD sum rule estimates on
$\gknl$ and $\gkns$\cite{ccl96,krippa98,choe98,bnn99,as00}, for  which
there are still ambiguities in among the  calculations.
Thus the results are quite different from each other.
More detailed analyses
are needed both experimentally and theoretically
to understand this discrepancy, and to understand
kaon-nuclear physics.

In Ref. \cite{ccl96,choe98}, the OPE was calculated only
up to the leading term coming from the
quark condensate and to leading order in $m_s$ in the sum rule
structure proportional to $\qsl i \gamma_5$.
However, the next leading term, dimension 5 $\mqc$ may contribute
to the OPE side with considerable amount as in
nucleon mass sum rule \cite{leinweber97}.
In addition, operators of dimension 7 may also be important in the
OPE side as a further power correction.
Thus, in this paper we re-analyze our QCD sum rule calculation
including higher dimensional condensates, such as
$\mqc$ and $\qc\gc$, and study the contribution of these condensates
on the previous results.

In Sec. \ref{sec2} we present our sum rules for $\gknl$ and $\gkns$,
and Sec. \ref{sec3} we discuss some uncertainties in our sum rules
and summarize our results.

%--------------------------------------------------------------------%
\section{QCD sum rules for $\gknl$ and $\gkns$ }
\label{sec2}

We will closely follow the procedures given in
Refs.\cite{rry83,rry85,ccl96,choe98}.
Consider the three point function constructed of the two baryon currents
 $\eta_B$, $\eta_{B'}$ and the pseudoscalar meson current $j_5$.
\beq
 \label{corr}
 A(p,p',q) = \int dx\, dy\,
 \la 0| T (\eta_{B'}(x) j_5(y) \overline{\eta}_B(0))|0 \ra \,
  e^{i(p'\cdot x - q\cdot y)} .
\eeq
In order to obtain $\gknl$, we will use the following
currents for the nucleon and the $\Lambda$ \cite{ioffe81,rry85}.
\beq
\eta_N&=&\epsilon_{abc}(u_a^T C\gamma_\mu u_b)\gamma_5 \gamma^\mu
d_c  ,
\label{etanuc}
\eeq
\beq
\eta_\Lambda&=&\sqrt{\frac{2}{3}} ~\epsilon_{abc} \left[
(u_a^T C\gamma_\mu s_b)\gamma_5\gamma^\mu d_c - (d_a^T C\gamma_\mu
s_b)\gamma_5\gamma^\mu u_c \right]  ,
\label{etalam}
\eeq
where u and d are the up and down quark fields ($a,b$ and $c$ are
color indices),
 $T$ denotes the transpose in Dirac space, and $C$ is the charge
conjugation matrix.
For the $K^-$ we choose the current
\beq
j_{K^-} = \bar{s}i\gamma_5 u  .
\eeq

The general expression for $A(p,p',q)$ has the following form
\beq
 A(p,p',q) &=& F_1 (p^2, p'^2, q^2) i\gamma_5 +
               F_2 (p^2, p'^2, q^2) \qsl i \gamma_5 \nonumber \\
           &+& F_3 (p^2, p'^2, q^2) \psl i\gamma_5 +
               F_4 (p^2, p'^2, q^2) \sigma^{\mu\nu} \gamma_5 q_\mu p'_\nu ,
\eeq
where $q = p' - p$ and $P = {p + p' \over 2}$.
Recently, in Ref. \cite{klo99} it was reported that in the case of
$g_{\pi NN}$ the
$ \sigma^{\mu\nu} \gamma_5 $ structure is independent of the
effective models employed in the phenomenological side and further
provides the $\pi NN$ coupling with less uncertainties from QCD
parameters.
Motivated by this result
$\gknl$ and $\gkns$ were calculated from  this structure
in Refs. \cite{bnn99,as00}.
In this paper, however, we construct the sum rule for only the
$\qsl i\gamma_5$ structure as before, and compare this with our previous
one.

On the phenomenological side, keeping the first two terms we have
\beq
&&\lambda_N \lambda_\Lambda
{M_B \over (p^2-M_N^2)(p'^2-M_\Lambda^2)} (\qsl i\gamma_5)
\gknl {1 \over q^2-m_K^2} {f_K m_K^2 \over 2 m_q}
\nonumber \\
&+& \lambda_{N} \lambda_{\Lambda^*}
{M'_B \over (p^2-M_N^2)(p'^2-M_{\Lambda^*}^2)} (\qsl i\gamma_5)
g_{KN\Lambda^*} {1 \over q^2-m_K^2} {f_K m_K^2 \over 2 m_q}
\nonumber \\
&+& \rm{higher ~resonances}  ,
\label{}
\eeq
where $M_B={1 \over 2} (M_N + M_\Lambda)$, and
$M'_B={1 \over 2} (M_N - M_\Lambda^*)$.
Here $\Lambda^*$ means the $\Lambda$(1405), and
we introduce (--) sign for the $\Lambda$ (1405) mass
because it is a negative parity state.
However, this is not relevant in the following calculation.
 $\lambda_N$, $\lambda_\Lambda$ and $\lambda_{\Lambda^*}$ are
the coupling strengths of the baryons
to their currents.
 $m_q$ is the average of the quark masses,
 $f_K$ the kaon decay constant and $m_K$ the kaon mass.
We take $f_K$ = 0.160 GeV and $m_s$ = 0.150 GeV.

As for the OPE side,
the new contribution from the quark-gluon condensates is given by
\beq
-\sqrt{2 \over 3}{7 \over 2^4 3 \pi^2}
\ln (-p^2) (\mqc + \msc)  ,
\eeq
and from dimension 7 ops.
\beq
+ \sqrt{2 \over 3}{5 \over 2^3 3^2}{1 \over p^2}(\qc + \s)\gc ,
\eeq
where we take the limit $p'^2 \rightarrow p^2$ and let $\uc = \dc
\equiv \qc$, $\muc = \mdc \equiv \mqc$. Here we collect only the
terms which contribute to the $\qsl/q^2$ structure such as Figs.
\ref{fig1} and \ref{fig2}. Using the standard values for $\msc =
0.8 ~\mqc$ and $\mqc = m_0^2 ~\qc = 0.8 ~\qc$ \cite{bi8283} the
sum rule after Borel transformation to $p^2=p'^2$ becomes
\beq
\lambda_N \lambda_\Lambda {M_B \over M_\Lambda^2-M_N^2}
\left(e^{-M_N^2/M^2} - e^{-M_\Lambda^2/M^2}\right) \gknl {f_K
m_K^2 \over 2 m_q} +
A ~\left(e^{-M_N^2/M^2} - e^{-M_{\Lambda^*}^2/M^2}\right) =
\nonumber\\
- ~\sqrt{2\over 3}
\left({33 \over 40\pi^2} E_1 M^4 +
({11m_s^2 \over 60\pi^2} - {21 \over 100 \pi^2}) E_0 M^2 +
~({m_s \over 3} \s + {1 \over 8} \gc) \right) \qc .
\label{sum1}
\eeq
Here,
A is the unknown constant coming from
$\lambda_{\Lambda^*} \cdot g_{KN \Lambda^*}$,
and
\beq
 E_i = 1 - \sum_{k=0}^i \frac{s_0^k}{k~!
~(M^2)^k} ~e^{-\frac{s_0}{M^2}} ,
\eeq
where $s_0$ is a continuum threshold.
One should be cautious, however, that there may be
non-accounted terms, which
can not be neglected by using this simple
Borel transformation\cite{ioffe95,kim99}.

For $\lambda_N$ and $\lambda_\Lambda$, we use the
values obtained from the
 following baryon sum rules for the $N$ and $\Lambda$ \cite{ioffe81,rry85}:
\beq
E_2^N M^6 + b E_0^N M^2 + {4 \over 3} a^2 = 2(2\pi)^4 \lambda_N^2
e^{-M_N^2/M^2} ,
\label{nucsum}
\eeq
\beq
E_2^\Lambda M^6 + {2 \over 3} a m_s (1 - 3\gamma) E_0^\Lambda M^2
+ b E_0^\Lambda M^2
+ {4 \over 9} a^2 (3 + 4\gamma)= 2(2\pi)^4 \lambda_\Lambda^2
e^{-M_\Lambda^2/M^2} ,
\label{lamsum}
\eeq
where $a \equiv -~(2\pi)^2 \qc $,
 $b \equiv \pi^2 \gc$,
and $\gamma \equiv \s / \qc -1 \simeq -~0.2$.
We use different thresholds for
$\lambda_N$ and $\lambda_\Lambda$ in
Eqs. (\ref{nucsum}) and (\ref{lamsum}).
We take $s_N$ = (1.440 GeV)$^2$ for the nucleon sum rule and
$s_\Lambda$ = (1.405 GeV)$^2$ for the $\Lambda$ sum rule considering
the next excited nucleon
and $\Lambda$ state, respectively.

$\gknl$ , however, does not display a plateau as
a function of the Borel mass. This is because there is no
usual power correction term like (${a \over M^2}, {b \over M^4}$, and so on)
in the r.h.s. of Eq. (\ref{sum1}) even including up to dimension 7 operators.
We need more higher dimensional operators to get those terms.
Thus, in this case we prefer to use
a best-fit method.
Eq. (\ref{sum1}) has the following form :
\beq
 \gknl \cdot f_1 (M^2) + A \cdot f_2 (M^2) = f_3 (M^2) .
\eeq
Then, we get $\gknl$ and the unknown constant A by  minimizing
$(\gknl \cdot f_1 + A \cdot f_2 - f_3)^2$ with a fixed $s_0$ and an
appropriate Borel interval:
\beq
\int_{M_{min}^2}^{M_{max}^2} (\gknl \cdot f_1 + A \cdot f_2 - f_3)^2 ~dM^2
= {\rm minimum} .
\eeq
We fix the continuum threshold $s_0$ = 2.074 GeV$^2$
taking into account the next term
from the N(1440), i.e., $N (1440) \rightarrow \Lambda$,
in the phenomenological side.

The Borel interval $M^2$ is restricted by the following conditions
: OPE convergence and pole dominance. The lower limit of $M^2$,
$M_{min}^2$ is determined as the value at which the contribution
of the highest dimensional operators is less than 10$\%$ of total
OPE. The upper limit $M_{max}^2$ is determined by restricting the
continuum contribution to be  less than 50$\%$. Then, we get
\beq |\gknl| &=& 2.49 ,
 \nonumber \\
 |A| &=& 0.00174\ {\rm GeV}^7,
\eeq
and the Borel interval (0.478, 1.068) GeV$^2$ for basic inputs
(i.e., $\qc$ = -- (0.230 GeV)$^3$, $\gc$ = 0.012 GeV$^4$, $m_s$ =
0.150 GeV, and $m_0^2$ = 0.8 GeV$^2$). Here we denote the absolute
value because we can not determine signs of the coupling strengths
($\lambda_N$, $\lambda_\Lambda$ and $\lambda_{\Lambda^*}$) in the
baryon sum rules. We also calculate the average deviation
$\bar{\delta}$ $\equiv$ $\sum_i^N \left| 1- RHS(M_i^2)/LHS(M_i^2)
\right| / N$ = 8.8 $\times$ 10$^{-2}$ to test the reliability of
our fitting, and it shows that the deviation is less than 10 \%.

Table \ref{table1} shows variations of $\gknl$ for
other input parameters,
which are coming from the uncertainty of the basic inputs.
 For example, the first line in Table \ref{table1} shows that
$|\gknl|$ = 3.11 (or 1.93) if we change the quark condensate to
$\qc$ = --(0.210 GeV)$^3$ (or --(0.250 GeV)$^3$) while other basic
inputs are fixed. In the last line we take $m_0^2$ = 0.6 GeV$^2$
from the lowest value of the standard QCD sum rule estimate
\cite{bi8283}, and 1.4 GeV$^2$ which was evaluated in the
instanton vacuum in Ref. \cite{pw96}. Total variation is about
$\pm$ 1.25 on the above $\gknl$ value. On the other hand, the
unknown constant $|A|$ varies from 0.00120 to 0.00203 GeV$^7$.

%--------------------------------------------------------------------%
Next, consider $\gkns$. The current of $\Sigma^\circ$ is obtained by
making an SU(3) rotation from the nucleon current\cite{cpw85}
\beq
\eta_\Sigma = \sqrt{2} ~\epsilon_{abc} \left[ (u_a^T C\gamma_\mu
s_b)\gamma_5\gamma^\mu d_c + (d_a^T C\gamma_\mu
s_b)\gamma_5\gamma^\mu u_c \right] .
\label{etasig}
\eeq
In this case the contribution of the quark-gluon condensate is
given by
\beq
-\sqrt{2} {1 \over 2^4 3 \pi^2} \ln (-p^2)
(\mqc + \msc) ,
\eeq
and from dimension 7 ops.
\beq
+ \sqrt{2}{1 \over 2^3 3^2}{1 \over p^2}(\qc + \s) \gc  .
\eeq
Then, within the same approximation as before we get the following
sum rule.
\beq
\lambda_N \lambda_\Sigma {M_B \over M_\Sigma^2-M_N^2}
\left(e^{-M_N^2/M^2} - e^{-M_\Sigma^2/M^2}\right) \gkns
{f_K m_K^2 \over 2 m_q} +
B \left(e^{-M_{N^*}^2/M^2} - e^{-M_\Sigma^2/M^2}\right) =
\nonumber\\*
+ ~\sqrt{2}
\left({3 \over 40\pi^2} E_1 M^4 +
({m_s^2 \over 60\pi^2} + {3 \over 100 \pi^2}) E_0 M^2
- {1 \over 40}\gc \right) \qc ,
\label{sum2}
\eeq
where $M_B = {1 \over 2} (M_N + M_\Sigma)$ and $N^*$ is N(1440).
B is the unknown constant coming from
$\lambda_{N^*} \cdot g_{KN^* \Sigma}$.
Again for  $\lambda_\Sigma$, we take the value
from the following sum rule for the $\Sigma$\cite{ioffe81,rry85}:
\beq E_2^\Sigma M^6 - 2 a m_s (1+\gamma) E_0^\Sigma M^2+b E_0^\Sigma M^2 +
{4 \over 3} a^2
=2(2\pi)^4 \lambda_\Sigma^2 e^{-M_\Sigma^2/M^2} .
\label{sigsum}
\eeq
We fix the continuum threshold $s_\Sigma$ = (1.660 GeV)$^2$ considering
the next $\Sigma$ state, $\Sigma$ (1660).

Using
the continuum threshold $s_0$ = 2.356 GeV$^2$ taking into account
the next term
from the N(1535), i.e., $N(1535) \rightarrow \Sigma$,
in the phenomenological side
we get
\beq |\gkns| &=& 0.395 ,
\nonumber \\
 |B| &=& 0.00148\ {\rm GeV}^7
\eeq
for the same basic inputs. The Borel interval is (0.488, 1.584)
GeV$^2$ and the average deviation of the fit $\bar{\delta}$ is 9.7
\% in this case. We present the variation of $\gkns$ on other
parameters in Table \ref{table2}. The total variation is about
$\pm$ 0.377. On the other hand, $|B|$ varies from 0.00117 to
0.00184 GeV$^7$.

%--------------------------------------------------------------------%
\section {Discussion}
\label{sec3}

SU(3) symmetry, using de Swart's convention\cite{swart63}, predicts
\beq
\gknl &=& - {1 \over \sqrt{3}} (3 - 2\alpha_D) g_{\pi NN} ,
 \nonumber \\*
 \gkns &=& + ~(2\alpha_D -1) g_{\pi NN}  ,
\label{su3}
\eeq
where  $\alpha_D$ is the fraction of the D type coupling,
 $\alpha_D = \frac{D}{D+F}$.
In Table \ref{table3} we compare our results with previous QCD sum
rule estimates \cite{krippa98,bnn99,as00} and an SU(3) symmetry
prediction, where we denote the error-bar allowing for SU(3)
symmetry breaking at the 20 \% level. Here we take $\alpha_D$ from
a recent analysis of hyperon semi-leptonic decay data by
Ratcliffe, $\alpha_D$=0.64\cite{ratcliffe96}, and $g_{\pi NN}$
from an analysis of the $np$ data by Ericson {\it et
al.}\cite{elbo96}, $g_{\pi NN}$=13.43. A comparison to fitting
analyses of experimental data\cite{trs95} is also provided. SU(3)
symmetry predicts $|\gknl/\gkns|$ = 3.55 taking $\alpha_D$ = 0.64,
while our results show that this ratio is 6.30 using the basic
inputs, and the order of SU(3) symmetry breaking is rather huge.

Let us remark on $g_{\pi NN}$ which was calculated in Ref.
\cite{rry83,rry85} using the three-point function method. After
including dimension 5 and 7 condensates as in the previous section
the sum rule becomes
\beq
\lambda_N^2 {e^{-M_N^2/M^2} \over M^2} M_N  g_{\pi NN}
{f_\pi m_\pi^2 \over \sqrt{2} m_q} +
C \left(e^{-M_{N^*}^2/M^2} - e^{-M_N^2/M^2} \right) =
\nonumber \\
- \left( {1 \over \pi^2} E_1 M^4 - {1 \over 5\pi^2} E_0 M^2
+ {1 \over 9} \gc \right) \qc ,
\label{sum3}
\eeq
where C is the unknown constant from $\lambda_{N^*} \cdot g_{\pi
NN^*}$ and $f_\pi$ = 0.133 GeV. The contribution of the
quark-gluon condensates in the OPE side is important as in the
$\gknl$ and $\gkns$ sum rules. In this case we use the PCAC
relation $f_\pi^2 m_\pi^2 = - 4 m_q \qc$ first, then the quark
condensate becomes an overall factor on both sides. However, the
coupling strength $\lambda_N$ is still related to the quark
condensate as shown in Eq. (\ref{nucsum}).

Following the same method in the previous section, and using $\qc$
= --(0.230 GeV)$^3$, $\gc$ = 0.012 GeV$^4$, and $s_0$ = 2.074
GeV$^2$ as a pure continuum threshold we get
\beq |g_{\pi NN}| &=& 3.65 \pm 2.31 , \nonumber \\ |C| &=& 0.00261
\pm 0.00091\ {\rm GeV}^7,
 \eeq
the Borel interval (0.460, 1.110) GeV$^2$, and the average
deviation of the fit $\bar{\delta}$ 9.3 \% at the central value.
Here the uncertainty comes from using different input parameters,
i.e. $\qc$ = --(0.210 GeV)$^3$ (or --(0.250 GeV)$^3$) , $\gc$ =
0.015 GeV$^4$, and $m_0^2$ = 0.6 (or 1.4 GeV$^2$) as before. In
this case the most error bar comes from uncertainties of the quark
condensate, i.e. from the coupling strength $\lambda_N$.

Now, let us discuss some uncertainties in our sum rules. In Eqs.
(\ref{sum1}), (\ref{sum2}) and (\ref{sum3}) the contribution of
the quark-gluon condensate is about 25 \%, 40 \%, and 20 \%,
respectively, of the leading term at $M^2$ = 1 GeV$^2$. Thus the
accurate value of this condensate is one of important factors in
our sum rules, and a more precise estimate may be needed (e.g.,
see Ref.\cite{op88}).

As we mentioned before, we need more higher dimensional operators
to get some power correction terms in our sum rules. Their
contribution will be much smaller than that of dimension 7
operators at the relevant Borel region around $M^2$ $\sim$ 1
GeV$^2$. However, those operators may contribute because the lower
limit of the Borel interval for each coupling constant is much
less than 1 GeV$^2$ in our sum rules.

We find that the coupling constants become 2 or 3 times larger
than the previous ones if we take the coupling strengths
($\lambda_N$, $\lambda_\Lambda$ and $\lambda_\Sigma$) from the
chiral-odd baryon sum rules \cite{ioffe81,rry85}. For example, we
get 7.04, 0.890, and 14.49 for $|\gknl|$, $|\gkns|$, and $|g_{\pi
NN}|$, respectively, for the basic inputs. Because the coupling
strengths from each baryon sum rule (the chiral-even and
chiral-odd) are not the same in the whole Borel region and the
discrepancy between the coupling strengths is larger in the low
Borel region, we get quite different coupling constants. Of
course, it should be judged by the stability of the sum rule
whether one chooses the coupling strengths from the chiral-even
sum rules or those from the chiral-odd sum rules.

As a final remark, in the case of $g_{\pi NN}$ it was shown that
there is a higher pseudoscalar resonance contamination from the
$\pi$ (1300) and $\pi$ (1800) in the three-point function method
\cite{maltman98}. Maybe there is a similar contamination from the
K(1460) and K(1830)\cite{prd98} on the kaon-baryon couplings.
Although the masses of the K(1460) and K(1830) are quite uncertain
and these states need further experimental confirmation, we can
briefly estimate the contribution of the K(1460) as done in Ref.
\cite{maltman98}. Using the parameters from recent works
\cite{vy99}, we get
\beq \left[ {f_M m_M^2 \over Q^2 + m_M^2} \right]_{Q^2 = 1\ {\rm
GeV}^2} = {\rm 21.3\ and\ 2.2\ MeV} \label{k1460} \eeq
for the kaon and K(1460), respectively.  Here $f_M$ is the decay
constant and $m_M$ is the meson mass. We take $f_K$ = 108 MeV,
$f_{K(1460)}$ = 3.3 MeV and $m_K$ = 496 MeV, $m_{K(1460)}$ = 1.45
GeV in Ref. \cite{vy99}. Comparing the values in Eq. (\ref{k1460})
to those for the pion and $\pi$ (1300)\cite{vy99}, i.e. 1.7 and
0.4 MeV, the contamination from the excited kaon state on the
kaon-baryon couplings seems smaller than that from the excited
pion state on $g_{\pi NN}$.

In summary, including higher dimensional condensates we re-analyze
our previous QCD sum rule estimate on $\gknl$ and $\gkns$ in the
$\qsl i \gamma_5$ structure. The contribution of dimension 5
quark-gluon condensates is comparable to that of the leading term,
and the present result is  much different from the previous one.

%
%------------------------ Acknowledgements --------------------------%
\acknowledgements

The author thanks Prof. Su H. Lee for valuable discussions and
comments. This work is supported by Research Fellowship of the
Japan Society for the Promotion of Science (JSPS).

%                                                                    %
%-----------------------  Table  ------------------------------------%
%--------------------------------------------------------------------%

\begin{table}
\caption{$\gknl$ and its variations. Other inputs mean other
possible inputs coming from the uncertainty of the basic inputs. }
 \label{table1}
\begin{center}
\begin{tabular}{ c c c}
 basic inputs &  other inputs  &  variations  \\
\hline
$ \qc $ = --(0.230 GeV)$^3$ & --(0.210 GeV)$^3$, --(0.250GeV)$^3$
                                                 & + 0.62, -- 0.53 \\
$ \gc $ = 0.012 GeV$^4$    & 0.015 GeV$^4$      &  + 0.45     \\
$m_s$ = 0.150 GeV   & 0.120, 0.180 GeV  &  -- 0.36, + 0.30    \\
$m_0^2$ = 0.8 GeV$^2$   & 0.6, 1.4 GeV$^2$ & -- 0.19, 0.02      \\
%\hline
\end{tabular}
\end{center}
\end{table}

\begin{table}
\caption{$\gkns$ and its variations. The same as in Table
\ref{table1}.}
\label{table2}
\begin{center}
\begin{tabular}{ c c c}
 basic inputs &  other inputs  &  variations \\
\hline
$ \qc $ =  --(0.230 GeV)$^3$ & --(0.210 GeV)$^3$, --(0.250 GeV)$^3$
                                                 &  + 0.057, -- 0.061 \\
$ \gc $ = 0.012 GeV$^4$     & 0.015 GeV$^4$      &  + 0.201     \\
$m_s$ = 0.150 GeV  & 0.120, 0.180 GeV  &  -- 0.084, + 0.086 \\
$m_0^2$ = 0.8 GeV$^2$   & 0.6, 1.4 GeV$^2$ & + 0.067, -- 0.198 \\
\end{tabular}
\end{center}
\end{table}

\begin{table}
\caption{Comparison of coupling constants.}
\label{table3}
\begin{center}
\begin{tabular}{c c c c c}
Sources  &             $g_{KN\Lambda}$   & $g_{KN\Sigma}$  \\
\hline SU(3) with 20 \% breaking  & -- 16.0 $\sim$ -- 10.7   & 3.0
$\sim$ 4.5 \\ Experimental fitting \cite{trs95}    & -- 13.7 & 3.9
\\ Ref. \cite{krippa98}  & 10 $\pm$ 6            &  3.6 $\pm$ 2 \\
Ref. \cite{bnn99}    & 2.37 $\pm$ 0.09    &   0.025 $\pm$ 0.015 \\
Ref. \cite{as00}   & 10 $\pm$ 2        & 0.75 $\pm$ 0.15   \\
Present work     &  2.49 $\pm$ 1.25      & 0.395 $\pm$ 0.377 \\
\end{tabular}
\end{center}
\end{table}

%
%------------------------ References --------------------------------%
%--------------------------------------------------------------------%

%
%-----------------Figure Captions-----------------------------------%
%-------------------------------------------------------------------%
\newpage
\begin{figure}
\caption{Contribution of dimension 5 operators. The solid lines are
quark propagators and the wavy line is a gluon propagator.
The dotted line denotes a meson.}

\label{fig1}
\end{figure}
\begin{figure}
\caption{Contribution of dimension 7 operators. The same as in
Fig. \ref{fig1}}
\label{fig2}
\end{figure}
%-------------------------------------------------------------------%
%
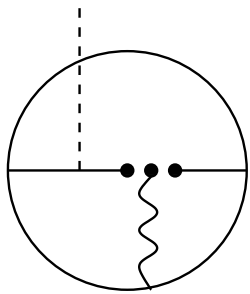
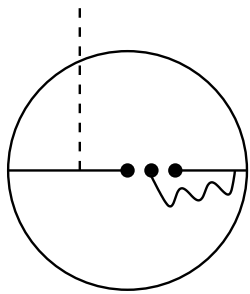
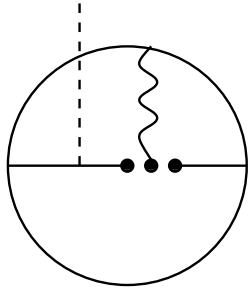
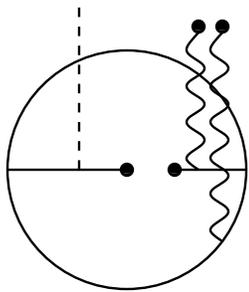
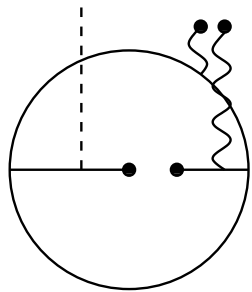
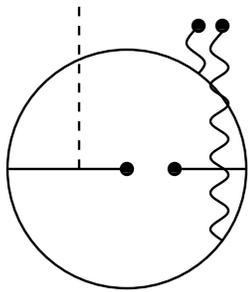
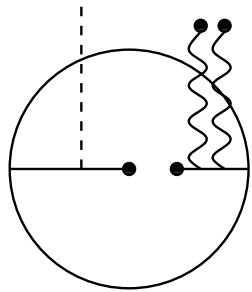
\end{document}